\documentclass{mpe_report}

\usepackage{psfig,graphicx,epsfig}
\usepackage{color}
\usepackage{amsmath,amssymb,epic,eepic,array}

\makeindex
\makeglossary

\begin{document}

%\pagenumbering{roman}
%
%\include{coverpage}  \clearpage
%\thispagestyle{empty}\phantom{x}\clearpage
%
%\include{coverpage2} \clearpage
%\thispagestyle{empty}\phantom{x}\clearpage
%
%\include{./Lesch/pref}  \clearpage
%\thispagestyle{empty}\phantom{x}\clearpage
%
%\include{participants}  \clearpage
%\thispagestyle{empty}\phantom{x}\clearpage
%
%\include{program} \clearpage
%
%
%\pagenumbering{Roman}
%\setcounter{page}{115}
%
%\include{tableofcontents} \clearpage
%

\pagenumbering{arabic}
\setcounter{page}{9}
\renewcommand{\FirstPageOfPaper }{  9}\renewcommand{\LastPageOfPaper }{ 12}%\documentclass{mpe_report}
%\usepackage{psfig}
% -----------------------------------------------------------------------------
%\def\R{~ROSAT}
%\def\RAS{\R all sky survey}

\newcommand{\rx}{RX J0822$-$4300}
\newcommand{\pu}{Puppis$-$A}
\newcommand{\RX}{RX~J0822$-$4300}
\newcommand{\Pu}{Puppis$-$A}

% -----------------------------------------------------------------------------
%\begin{document}

\title{A Fast-Moving Central Compact Object in Puppis-A}
\author{C. Y. Hui \and W. Becker}  
\institute{Max--Planck--Institut f\"ur extraterrestrische Physik,
 Giessenbachstra{\ss}e, 85740 Garching, Germany}
\maketitle

\begin{abstract}
Utilizing two Chandra High Resolution Camera (HRC-I) observations with an epoch separation of 
somewhat more than five years, we have measured the proper motion of the central compact object, 
RX J0822-4300, in the supernova remnant Puppis-A for the first time. The position of 
RX J0822-4300 is found to be different by $0.574\pm0.184$ arcsec, implying a proper motion 
of $107.49\pm34.46$ mas/yr with a position angle of $241^{\circ}\pm24^{\circ}$. For a
distance of 2.2 kpc, this proper motion is equivalent to a recoil velocity of 
$1121.79\pm359.60$ km/s. Both the magnitude and the direction of the proper motion are in 
agreement with the birth place of RX J0822-4300 being near to the optical expansion center 
of the supernova remnant. Although the positional shift inferred from the current data is 
significant at a $\sim3\sigma$ level only, one or more future HRC-I observations can obtain a 
much larger positional separation and further constrain the measurement. 
\end{abstract}

\section{Introduction}
Thanks to the much improved sensitivity of the state-of-art
X-ray observatories, the group of supernova remnants (SNRs) which are known
to host central compact objects (CCOs) is a slowly growing one (see Hui \&
Becker 2006a and reference therein). CCOs are generally characterized by
their high X-ray to optical/radio flux ratio, the lack of long-term
variability as well as their locations near to the expansion centers of SNRs.
One of the most interesting CCO is RX J0822-4300 located in the SNR Puppis-A.

Recently, we have presented results from a detailed study of
RX J0822-4300 which made use of all XMM-Newton and Chandra avaliable
by now (Hui \& Becker 2006a, 2006b). The spectral analysis of XMM-Newton
data revealed that the X-ray emission from RX J0822-4300 is in
agreement with being of thermal origin. A double blackbody model
with the temperatures $T_{1}=(2.35-2.91)\times10^{6}$ K,
$T_{2}=(4.84-5.3)\times10^{6}$ K and the projected blackbody emitting
radii $R_{1}=(2.55-4.41)$ km, $R_{2}=(600-870)$ m gave the best
description. Though a blackbody plus a very steep power-law tail
($\alpha>3$) yields a compatible goodness-of-fit relative to that of the
double blackbody model (cf.~Table 2 in Hui \& Becker 2006b).

The X-ray images taken with the Chandra HRC-I camera allowed for the first time to
examine the spatial nature of \RX\, with sub-arcsecond resolution. Besides an accurate
measurement of the source position, this observation constrained the point
source nature of \rx\, down to $0.59\pm0.01$ arcsec (FWHM) for the first time.

Despite the effort in searching for coherent radio pulsations, \rx\, has not been detected
as a radio pulsar (Kaspi et al. 1996). Similar to many other CCOs, it has no optical
counterpart down to a limiting magnitude of $B\ga 25.0$ and $R\ga 23.6$ (Petre, Becker,
\& Winkler 1996). Together with the lack of long term flux variation (Hui \& Becker 2006b),
all these evidences rule out many types of X-ray sources as a likely counterpart of
\rx, except a neutron star.

\rx\, is located about $\sim 6$ arcmin distant from the optical expansion center of \pu,
which is at RA=$08^{\rm h}22^{\rm m}27.45^{\rm s}$ and Dec=$-42^{\circ}57'28.6"$ (J2000)
(cf.~Winkler \& Kirshner 1985; Winkler et al.~1988). The age of the supernova remnant,
as estimated from the kinematics of oxygen-rich filaments is $\sim 3700$ years (Winkler
et al.~1988). If these estimates are correct, \rx\, should have a proper motion of
the order of $\sim 100$ mas/yr to a direction away from its proposed birth place.

We have tested this hypothesis by making use of archival Chandra HRC-I data
spanning an epoch of somewhat more than five years (Hui \& Becker 2006a). 
The expected positional displacement
for \RX\ in this time span is of the order of $\sim 0.5$ arcsec. This is in the range of
the Chandra accuracy given the possibility to correct for pointing uncertainties by using
X-ray counterparts of stars with accurate position and which are serendipitously located
in the field of view.

\section{Observation and data analysis}
Owing to the fact that there are only few bad pixels in the Chandra HRC-I and the pixel
size of 0.13187 arcsec oversamples the point spread function (PSF) by a factor of $\sim 5$,
the HRC-I appears to be the most suitable detector to perform astrometric measurements of
X-ray sources.

Checking the Chandra archive for suitable data we found that by mid of 2006 \rx\, was
observed twice with the HRC-I. The first observation was performed in 1999 December 21
(MJD 51533) for an exposure time of about 16 ksec. The second observation was done in
2005 April 25 (MJD 53485) for an on-source time of $\sim 33$ ksec. In the April 2005
observation, the target was displaced only $\sim 0.2$ arcmin off from the optical axis of
the X-ray telescope. In the December 1999 data, it appears to be $\sim 0.3$ arcmin
off-axis which, in both cases, is small enough to have a negligible effect for the
distortion of the PSF relative to an on-axis observation.

In order to increase the precision required for accurate astrometric measurements,
systematic uncertainties need to be corrected. Apart from the aspect offset
correction we also considered the errors introduced by determining the event positions.
The later included corrections of the tap-ringing distortion in the position reconstruction
and the correction of errors introduced by determining the centroid of the charge cloud.
Instead of using the fully processed pipeline products, we started our data reduction
with level-1 files to be able to correct for these systematic effects. All the data
processings were performed with CIAO 3.2.1. Details of the applied corrections are
described in the following.

Instabilities in the HRC electronics can lead to a tap-ringing distortion in the
position reconstruction of events. Correction has been applied to minimize this
distortions in standard HRC level-1 processing which required to know the values
of the amplitude scale factor (AMP\_SF). Such values are found in the HRC telemetry and
are different for each event. Unfortunately, they are often telemetered incorrectly.
In order to fix this anomaly, we followed a thread in CIAO to deduce the correct
values of AMP\_SF in the level-1 event file from other HRC event data and applied
these corrected values to minimize the distortion.

The de-gap correction was applied to the event files in order to compensate
the systematic errors introduced in the event positions by the algorithm which
determines the centroid of the charge cloud exiting the HRC rear micro-channel
plate.

After correcting these systematic errors we generated the level-2 event lists
files which were used thoroughly for the remaining analysis. We created HRC-I
images of \rx\, for both epochs with a binning factor of 1 so that each pixel
has a side length of 0.13187 arcsec.

To be able to correct for pointing uncertainties by using X-ray counterparts of
stars which have their position known with high accuracy, we applied a wavelet
source detection algorithm to the HRC-I images. Two X-ray point sources with
a count rate of about 1\% and 3\% relative to that of \rx\, were detected
serendipitously at about $\sim2.5$ arcmin and $\sim5.5$ arcmin distance from \rx. Figure 1
shows a $9.5\times 7$ arcmin field surrounding \rx\, as seen with the HRC-I in April 2005.
Both serendipitous sources, denoted as A and B, are indicated in this figure.
Their X-ray properties are summarized in Table 1.

\begin{table*}
\caption{X-ray properties of serendipitous sources in the neighborhood of \rx.}
\centering
\begin{tabular}{lcccc}
\hline\hline
Source & \multicolumn{2}{c}{X-ray position} & Positional error$^a$ & Net count rate \\
        & RA (J2000) & Dec (J2000) &   & cts/s \\
\hline\\[-1.5ex]
\multicolumn{5}{c}{{\bf 1999 observation}}\\[1ex]
\hline
A & $08^{\rm h}21^{\rm m}46.339^{\rm s}$ & $-43^{\circ}02'03.73"$ & 0.16" & $(2.51\pm 0.38)\times 10^{-3}$ \\
B & $08^{\rm h}22^{\rm m}23.924^{\rm s}$ & $-42^{\circ}57'59.29"$ & 0.20" & $(8.42\pm 0.75)\times 10^{-3}$ \\
\hline\\[-1.5ex]
\multicolumn{5}{c}{{\bf 2005 observation}}\\[1ex]
\hline
A & $08^{\rm h}21^{\rm m}46.313^{\rm s}$ & $-43^{\circ}02'03.46"$ & 0.07" & $(3.20\pm 0.30)\times 10^{-3}$\\
B & $08^{\rm h}22^{\rm m}23.966^{\rm s}$ & $-42^{\circ}57'59.70"$ & 0.18" & $(7.47\pm 0.48)\times 10^{-3}$ \\
\hline
\end{tabular}
\\
$^a$ The positional errors of sources A and B are given by the PSF fitting and the wavelet detection algorithm
respectively.
\end{table*}

In order to determine their X-ray positions with higher accuracy than possible
with the wavelet analysis, we fitted a 2-D Gaussian model with the modeled PSF
as a convolution kernel to both sources A and B. These fits require some
information on the source energy spectrum which is not available from the HRC-I
data. We therefore checked the archival XMM-Newton data for both sources and
found from an spectro-imaging analysis of MOS1/2 data (cf.~Figure 1 in Hui \& Becker
2006b) that the hardness-ratios of source A and B are comparable to that of \rx\,
which has its energy peak at $\sim 1.5$ keV. We therefore extract the
Chandra HRC-I PSF model images at 1.5 keV with the corresponding off-axis angles
from the CALDB standard library files
(F1) by interpolating within the energy and off-axis grids using the CIAO tool
MKPSF. The exposure maps were also generated at this energy by using MKEXPMAP.

The size of the $1-\sigma$ error circles of source A obtained by this method are
0.16 arcsec and 0.07 arcsec for the 1999 and 2005 observations, respectively.
The relatively large error in the December 1999 observation is due to its shorter
integration time and thus smaller photon statistics.

The large off-axis angle ($\sim 5.5$ arcmin) of source B causes a marked blurring
of the PSF (90\% encircled energy radius $\simeq 4$ arcsec). Such distortion makes
the source appear to be very dispersed. Given the patchy and uneven supernova
remnant emission this source is surrounded by and the limited photon statistics
we did not succeed in obtaining its coordinates more accurate than possible with the
wavelet algorithm (which is $\sim 0.2$ arcsec). This leaves source A as the only
reference star to perform astrometric correction.

Correlating the source position of source A with the Two Micron All Sky Survey (2MASS) catalog
(Skrutskie et al.~2006) identified the star with the source designation 08214628$-$4302037
as a possible optical counterpart. Since the next nearest optical source is located
about 5 arcsec away from the X-ray position of source A, we adopt the 2MASS source
08214628$-$4302037 as its optical counterpart. The visual magnitudes of the object
are $J=12.161\pm 0.027$, $H=11.675\pm 0.023$ and $K=11.558\pm 0.024$.
Since its spectral type is not known with certainty, we adopted a typical X-ray-to-optical
flux ratio of log$(F_x/F_{opt})\simeq -2.46$ for stars from  Krautter et al.~(1999).
Assuming a Raymond-Smith thermal plasma model with $kT=0.15$ keV, $n_H=4\times10^{21}$
cm$^{-2}$ (Hui \& Becker 2006b) and solar abundances for the star's spectrum we estimated
with the aid of PIMMS (version 3.8a2) its HRC-I count rate to be $\sim 3\times10^{-3}$
cts/s. This is in good agreement with the observed count rate of source A (cf.~Table 1).

In order to use the optical identification of the serendipitous X-ray source A as a reference source
for the offset correction, we have to check whether itself shows a proper
motion. To investigate this, we correlated its 2MASS position with the UCAC2 catalog
(Zacharias et al. 2003). We unambiguously found a source with the UCAC2 designation 13302738
as a counterpart of the X-ray source A. According to this catalog, this source has a proper motion
of $\mu_{\rm RA}=-16.0\pm5.2$ mas/yr, $\mu_{\rm dec}=-1.7\pm5.2$ mas/yr.

We attempted to make an independent estimate by
analysing the images from the first and the second
Digitized Sky Surveys\footnote{http://ledas-www.star.le.ac.uk/DSSimage/}.
From the observation dates specified for the DSS-1 and DSS-2 images,
we found that the epoches of these two images are separated by 5134 days.
We took four bright stars within 1 arcmin neighborhood of the X-ray source A
as the references to align the frames of DSS-1 and DSS-2.
None of these stars appeared to be saturated so that their
positions could be properly determined by a 2-D Gaussian fit.
We determined the offset between these two images from comparing the
best-fit positions of the four reference stars in both frames. However, we
found that the alignment error is at a level of $\sim0.5$ arcsec. This is
close to the average positional discrepancy between DSS-1 and DSS-2 which
is found to be $\sim0.6$ arcsec. The information provided by DSS-1/2
thus does not allow us to estimate the proper motion of our source of interest
independently.  In view of this, we can only resort on the findings in the
UCAC2 catalog.

Under the assumption that the UCAC2 object 13302738
is indeed the optical counterpart of source A, we applied the aspect
correction to the corresponding frames with the proper motion of the reference star
taken into account. However, with only one comparison
source available for the frame alignment, there are some limitations in our
adopted method. Firstly, the roll angle between two frames cannot be
determined independently with just one reference source. Hence, the accuracy
of the current result is limited by the output of the star-tracking camera, the
Pointing Control and Determination system (PCAD). Also, an independent
estimate of the HRC-I plate scale cannot be made with only one reference
source. The potential variation of the plate scale might introduce an extra
error, though we consider this is negligible as the typical uncertainty of
the HRC-I plate scale is at the order of $\sim0.05$ mas/pixel
\footnote{\scriptsize{http://cxc.harvard.edu/cal/Hrma/optaxis/platescale/geom\_public.html}}.

The error circle of the UCAC2 object 13302738 is specified to be 0.015 arcsec. Including the uncertainty in the
proper motion, the overall positional error of this astrometric source is increased with time which gives
0.016 arcsec and 0.037 arcsec in the December 1999 and April 2005 epoch respectively.
The total error of the aspect corrections is calculated by combining the statistical error of the X-ray position of
source A and the astrometric error of the UCAC2 object in quadrature. This yields a $1-\sigma$
error of 0.161 arcsec and 0.079 arcsec for the aspect correction of the December 1999 and April 2005
observation, respectively.

The position of \rx\, was determined by the same procedure we applied to obtain
the position for source A. The fits provide
us with the coordinates for \rx\, which are RA=$08^{\rm h}21^{\rm m}57.389^{\rm s}$
and Dec=$-43^{\circ}00'16.90"$ (J2000) in the 1999 observation and
RA=$08^{\rm h}21^{\rm m}57.343^{\rm s}$ and Dec=$-43^{\circ}00'17.18"$ (J2000)
in the April 2005 observation\footnote{The best-fitting position in the 1999 observation
has a 0.242 arcsec deviation from that inferred by Hui \& Becker (2006b) which
corrected the offset by simply using the calculator provided by the Chandra Aspect team
http://cxc.harvard.edu/cal/ASPECT/fix\_offset/fix\_offset.cgi}.
The  $1-\sigma$ error of the position introduced by the PSF-fit is
0.01 arcsec in both epochs.
In order to exclude any dependence of the deduced source positions on the aperture size
of the selected source region, we repeated the fits for three different apertures with
radii equal to 3 arcsec, 4 arcsec and 5 arcsec, respectively. We did not find any
variation of the best-fitted parameters in these independent fittings.

\begin{figure}
\psfig{figure=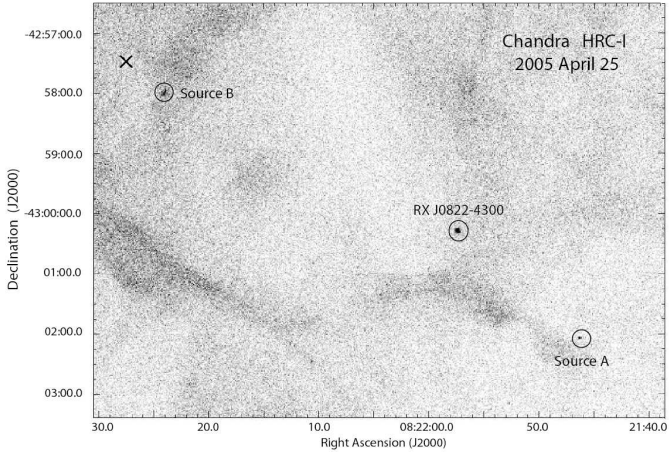,width=9cm,clip=}
\caption{Part of the Chandra HRC-I image of the \pu\ supernova remnant. Besides \rx, there are
 two serendipitous point sources detected in the field of view. They are denoted as source A and
  B. The optical expansion center of \pu\ as obtained by Winkler et al.~(1988) is marked by a cross.
  }
\end{figure}

\section{Results and discussion}
The fitted X-ray positions of \rx\, are plotted for the observational epochs
December 1999 and April 2005 in Figure 2. The size of the error circle for each position
is determined by adding the error in correcting the aspect offset and the error of the
position of \rx\, in quadrature. This gives 0.162 arcsec and 0.088 arcsec for the December
1999 and the April 2005 observations, respectively. The corresponding error circles are
indicated in Figure 2.

Comparing the positions of \rx\, as deduced for the two observations in 1999 and 2005
we found that they are different by $0.574\pm 0.184$ arcsec. This is well consistent with the displacement
estimated from the kinematic age of the SNR and the positional offset of \rx\, from the
SNR's optical expansion center (cf.~\S1). The quoted error is $1-\sigma$ and combines
the positional errors in both epochs in quadrature.

Given the epoch separation of 5.34 years for both Chandra HRC-I observations the observed
proper motion of \rx\, is $\mu=107.49\pm 34.46$ mas/yr. The position angle is PA=$241^{\circ}\pm24^{\circ}$.
Taking the position inferred from the December 1999 observation and the position of the
optical expansion center of \pu\, (Winkler et al.~1988), a position angle of
PA$\simeq 243^{\circ}$ is implied. Such consistency suggests that \rx\, is indeed physically
associated with \pu\, and its actual birth place is not too far away from the SNR's optical
expansion center. Assuming that the distance of 2.2 kpc to \pu\, is correct, the observed proper motion
implies that \rx\, has a projected recoil velocity as high as $1121.79\pm 359.60$ km/s to the
southwest. 

Although our result is a promising indication of a fast moving CCO in a SNR,  we note that the
deduced positional shift is only significant at a $3.1\sigma$ level.
Since the X-ray source A is rather faint, its relatively large positional error predominates in the error budget.
In order to further constrain this measurement, one or more Chandra
HRC-I observations can attain a much larger positional separation in a few years from today.

\begin{figure}
\centerline{\psfig{figure=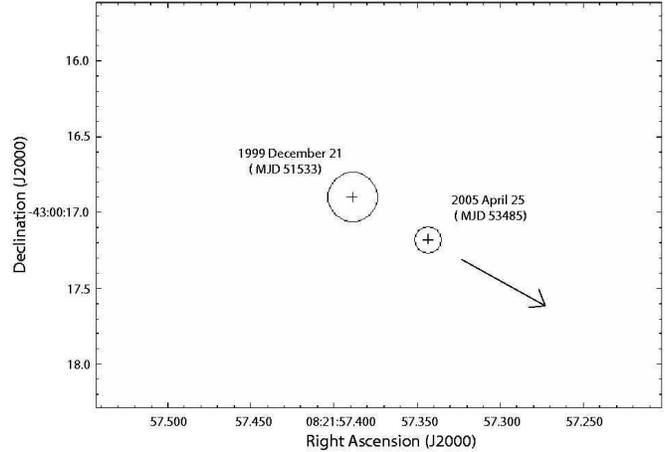,width=9cm,clip=}}
\caption{The best-fitted X-ray positions of \rx\, in two epochs are marked by crosses.
The circles indicate the $1-\sigma$ error.
The arrow shows the direction of proper motion inferred from both positions.
	      }
\end{figure}

After the letter of Hui \& Becker (2006a) was accepted for publication in A\&A, 
we became aware that a paper on the similar subject as presented here was
submitted to ApJ by Winkler \& Petre (2006). In their paper, though using the same dataset as we use,
a much higher significance ($9.4\sigma$ !) for the proper motion of \rx\ is claimed.
They justify this by their inclusion of source B as a second reference star.
In view of its large off-axis angle, thus blurred PSF and large error circle ($\sim0.2$ arcsec), we question their
justification and doubt their error estimate being a realistic value, even their proper motion estimate is qualitatively
similar to ours.

%\end{document}

                \clearpage

\end{document}